\documentclass{svjour3}                     
\smartqed  

\usepackage[numbers]{natbib}
\usepackage{graphicx}

\begin{document}

\title{Examining the Kerr Metric through Wave Fronts of Null Geodesics}

\author{Thomas P. Kling         \and
        Eric Grotzke           \and
        Kevin Roebuck            \and
        Harry Waite
}


\institute{T. Kling \at
              Dept. of Physics, Bridgewater State University,
Bridgewater, MA 02325 \\
              Tel.: +1-508-531-2895\\
              Fax: +1-508-531-1785\\
              \email{tkling@bridgew.edu}           
           \and
            E. Grotzke \at
             Dept. of Physics, Wake Forest University, Winston-Salem,  NC 27109
            \and
           K. Roebuck \at
              Dept. of Physics, Wake Forest University, Winston-Salem,  NC 27109
             \and
             H. Waite \at
             Dept. of Physics, Bridgewater State University,
Bridgewater, MA 02325
}

\date{Received: date / Accepted: date}

\maketitle

\begin{abstract}

\noindent We examine the singularities of the wave fronts of null geodesics from point sources in the Kerr Metric.  We find that the wave fronts develop a tube like structure that collapses non-symmetrically, leading to cusp features in the wave front singularities.  As the wave front advances, the cusps trace out an astroidal shaped caustic tube, which had been discovered previously using lens mapping and geodesic deviation methods.  Thus, the wave front approach in this study helps to complete a picture of caustics and gravitational lensing in the Kerr geometry.

\end{abstract}


\maketitle


\section{Introduction} \label{intro:sec}

The Kerr metric, representing the space-time surrounding a rotating black hole, is likely the most important of the stationary solutions to the Einstein Field Equations.  The metric played an important historical role in the development of advanced techniques of general relativity, in particular for asymptotically flat space-times \cite{NP}.  Because it is the asymptotically flat solution of that can represent the exterior of a isolated, rotating body, its properties are particularly important in a range of applications, including solar system tests of general relativity such as Gravity Probe B \cite{everitt} and examinations of the black hole at the center of our galaxy including ideas such as the black hole shadow \cite{mizuno}.

An examination of wave fronts of light rays provides an alternative, and complementary, way to examine the geometrical approach to gravitational lensing \cite{ehlers, fp, pettersbook}.  Studies of gravitational lensing by Schwarzschild black holes were initiated by Virbhadra and Ellis (2000) and Frittelli et al. (2000) \cite{virbhadra, frittelli}, and the subject in general has been well reviewed and results summarized by Perlick \cite{perlick}.

Kerr black hole lensing followed studies of lensing by Schwarzschild black holes.  A lens equation, showing how the images of objects behind the lens would be viewed by an observer, was derived by Bozza (2003) \cite{bozza02}, and similar work was developed independently by Sereno (2003) \cite{sereno2003}. A more recent set of lens equations was introduced by Aazami et al. (2011) \cite{aazami}, and photon trajectories in extreme Kerr black holes ($a>m$) were considered by Porfyiadis et. al (2016) \cite{porfy}.   These papers highlighted some of the differences in lensing properties with the addition of rotation.

An important paper by Rauch and Blandford (1992) computed light rays around Kerr black holes and considered the caustics of the light rays \cite{rauch}.  This paper used a combination of analytic work, integral expressions for the light rays, and what was, for the time, stunning computational work to visualize an astroidal shaped  caustic tube that circled the Kerr black hole.  The work of Rauch and Blandford to demonstrate the astroidal caustics was rederived in a series of recent papers using geometrical and lensing methods \cite{sereno2008, bozza2008, grould}.

The current study re-examines Kerr black hole lensing with a focus on showing the wave fronts of point sources in the space-time.  These wave fronts provide a complementary picture to the analytic work of Bozza et al. (2005) and Aazami et al. (2011) on Kerr black hole lensing \cite{bozza05, aazami} and show how the caustic tube of Rauch and Blandford arises \cite{rauch}.  Our paper uses highly accurate numerical integration of the null geodesic equations and custom visualization software to create movies of the wave fronts.  Our techniques are similar to those of Grould et al. (2008)\cite{grould} in that we numerically integrate the geodesic equations, although we use different coordinates to achieve better properties in error correction.  Other attempts to demonstrate wave fronts in the Kerr geometry reveal some aspects of the key features, but at lower resolutions that do not allow one to see how the caustics arise \cite{frutos}.  To our knowledge, this is the first paper that shows how the astroidal caustics arise from the wave front perspective.

In Section \ref{equations:sec} we outline previous approaches to integrating the null geodesics in the Kerr space-time.   Section~\ref{numerical:sec} outlines our numerical approach in contrast to the more analytic and quasi-analytic approaches taken previously including discussions of the boundary conditions necessary to achieve a sphere's worth of null geodesics that are past directed. We examine properties of individual light rays and discuss details of the numerical resolution achieved in Section~\ref{rays:sec}.  The origin of astroidal caustics is explained in Section~\ref{astroidal:sec} and it is shown why these do not develop in the Schwarzschild case in Section~\ref{schwarz:sec}. Kerr wave fronts are discussed for both point sources in the azimuthal plane in Section~\ref{wavefronts:sec} and briefly for initial points outside the azimuthal plane in Section~\ref{outofplane:sec}.  We discuss the relationship between the astroidal casustics arising along wave fronts and the caustics of the lens mapping in Section~\ref{discussion:sec}.  In an appendix, we discuss issues related to future null geodesics in our numerical approach.


\section{Traditional Approach to Null Geodesics of Kerr Metric} \label{equations:sec}

In the traditional presentation of the Kerr metric given by Hawking and Ellis \cite{HE}, Boyer and Lindquist coordinates are used, and the Kerr metric is given by

\begin{equation}  ds^2 = \rho^2 \left( \frac{dr^2}{\Delta} + d\theta^2 \right) + (r^2 + a^2) \sin^2 \theta d\phi^2 - dt^2 + \frac{2mr}{\rho^2} (a\sin^2\theta d\phi - dt)^2 \label{kerr1}, \end{equation}

\noindent where the functions $\rho$ and $\Delta$ are defined as

\begin{eqnarray} \rho^2(r, \theta) &= & r^2 + a^2 \cos^2 \theta \\ \Delta & = & r^2 - 2mr + a^2. \end{eqnarray}

\noindent In the metric, $m$ represents the mass of the black hole and $a$ is the spin parameter with $ma$ the angular momentum.  The metric reduces to the Schwarzschild solution when $a=0$.

Properties of light rays in Kerr metrics date back to early work by Cunningham and Bardeen (1972) \cite{cunningham}.  In fact, quite a lot is known about the null geodesics of the Kerr metric using first integrals of the motion \cite{chandrasekhar}.   The seminal paper by Rauch and Blandford reduces the geodesic equations to elliptical integrals which are then integrated numerically \cite{rauch}.  Bozza (2003) used four first integrals of the motion to derive first order ordinary differential equations for null geodesics \cite{bozza02}.  Approximation schemes were then utilized to look at light rays connecting sources and observers far from the black hole.  A later Kerr black hole lensing paper in the strong field regime
 utilized the equations for the null geodesics in an integral formulation introduced by Carter (1968) \cite{carter, bozza05}.

While all of these approaches obtain reasonably simple equations for individual null geodesics, they involve features that complicate efforts to solve for the large numbers of null geodesics simultaneously needed to construct a wave front and examine wave front singularities.  In particular, the versions involving elliptic integrals used by Rauch and Blandford and most of the later papers used Gauss-Kronrod methods that are designed to compute an integral between a starting $r_i$ and a final $r_f$, or initial and final $\theta$ coordinates, in one accurate step.   The integral approaches are particularly poorly suited to watching the wave fronts move, which helps one understand how the wave front singularities arise. Further, the methods that utilize first-integrals of the motion typically require finding a minimum distance of approach where expressions in the null geodesic equations change sign, a computationally heavy process when integrating many rays.

\section{Numerical Approach to Null Geodesics for Kerr Wave Fronts} \label{numerical:sec}

In this paper, we will be simultaneously integrating hundreds of thousands of null geodesics of the Kerr metric. Our choice of coordinates and the way we write the geodesic equations for the Kerr metric are optimized to allow us to obtain highly accurate tracings of many light rays simultaneously, and to re-organize the points along all of the light rays to find spatial positions at given times. Traditional approaches to the null geodesic equations are not well suited to the approach needed to create highly accurate wave fronts.  Integral methods are well suited to finding one final location along a ray, but not continuously integrating out each ray.  Further, the first integrals often used result in equations that do not align well with the process of finding wave fronts because they do not provide a natural way to determine where the light ray is in space at a specified time.

Numerically, our implementation will involve a modification of the Runge-Kutta-Fehlberg 4-5 adaptive step-size approach to simultaneously integrate the geodesic equations as ordinary different equations \cite{nrc}, similar to the approach of Grould et al (2016) \cite{grould}.  Adaptive step-size approaches allow one to monitor the accumulated error in the null geodesics and ensure highly accurate steps even near the Kerr event horizon.

We are ultimately interested in wave fronts, with each null geodesic advancing the local region of the wave front. We locate each null geodesic's contribution to the $t=t_1$ wave front by the geodesic's position when in a bin $t_1 + \epsilon$ for small $\epsilon$.  If more than one time step is within the same time bin, we take only the first time step.  Our three-dimensional visualizations are accomplished by separate code that receives data points from the adaptive step-size numerical integration program and then sorts and orders the data according to time. This allows us to examine the advancing of the moving wave fronts in time to better understand how the wave fronts fold and develop wave front singularities.  Our custom code allows us to continuously pan, change the vantage point, rotate the image, and zoom in and out on the wave front, which are critical tools because the projection of wave fronts into a flat plane can introduce false singularities that vanish under small changes in vantage point.  True wave front singularities are stable under these changes.

\subsection{Euler-Lagrange Equations for Wave Fronts in Kerr Coordinates}

Examining the metric in Eq.~\ref{kerr1}, we see that the $\Delta$ in the denominator of the $dr^2$ term vanishes at the event horizon causing a coordinate singularity in the metric.  Because of this, the null geodesic equations have singular terms in the $(t, r, \theta, \phi)$ coordinates.  As a result, adaptive step-size methods fail as one approaches the Kerr event horizon because the infinity in the equations drives the step-size to zero.

Therefore, we will instead integrate the null geodesic equations in the Kerr coordinates $(r,\theta, \phi_-, u_-)$ where one can extend the metric across the event horizon \cite{HE} for past-directed null geodesics.  (We discuss what happens to our adaptive step-size code if we use the $(r, \theta, \phi_-, u_-)$ coordinates for future-directed geodesics, as well as the appropriate choice of $(r, \theta, \phi_+, u_+)$ coordinates for that time orientation in the Appendix.)  Our choice of coordinates differs from other approaches to treating the geoedesic equations as ODEs \cite{grould, frutos}, and allows us to obtain high accuracy with fewer steps close to the black hole.  This in turn allows us to simultaneously integrate more rays along the wave front because less computational time is required by the adaptive step-size algorithm in this choice of coordinates.

The $u_-$ and $\phi_-$ coordinates are defined by the relations

\begin{eqnarray}  du_- &=& dt - \frac{(r^2 + a^2)}{\Delta} dr \label{uminusdef} \\ d\phi_- &= & d\phi - \frac{a}{\Delta} dr. \label{phiminusdef}\end{eqnarray}

\noindent In these coordinates, the metric is given by

\begin{eqnarray} ds^2  & = & \rho^2 d\theta^2   +  2 a \sin^2\theta dr d\phi_- - 2 dr du_- + \rho^{-2}[(r^2+a^2)^2 - \Delta a^2 \sin^2 \theta] \sin^2\theta d\phi_-^2 \nonumber \\ & ~ & ~ -4a\rho^{-2} m r \sin^2\theta d\phi_- du_- - (1-2mr\rho^{-2}) du_-^2 . \label{kerr2} \end{eqnarray}

Collecting terms in front of each coordinate variation as functions written as $T_i$, we define a Lagrangian for Kerr null geodesics as

\begin{eqnarray} {\mathcal{L}}  & = & \frac{1}{2} g_{ab} \dot x^a \dot x^b  \nonumber \\ &=& T_1 \dot \theta^2 + T_2 \dot r \dot \phi_- + T_3 \dot r \dot u_- + T_4 \dot \phi_-^2 + T_5 \dot \phi_- \dot u_- + T_6 \dot u_-^2 , \label{lagrangian} \end{eqnarray}

\noindent  where the dot represents a derivative with respect to an affine parameter $s$.  The explicit form of the $T_i$ are given by

\begin{eqnarray} T_1 & = & \frac{\rho^2}{2} \nonumber \\
T_2 & = &  a \sin^2\theta \nonumber \\
T_3 & = &  -1 \nonumber \\
T_4 & = & \frac{1}{2\rho^2} [(r^2+a^2)^2 - \Delta a^2 \sin^2 \theta] \sin^2\theta  \nonumber \\
T_5 & = & - 2 a\rho^{-2} m r \sin^2\theta \nonumber \\
T_6 & = & - \frac{1}{2} (1-2mr\rho^{-2}). \nonumber \end{eqnarray}

The $u_-$ and $\phi_-$ coordinates are cyclic, which implies the existence of two constants of integration that would allow one to write the Euler-Lagrange equations as two second-order and two first-order ordinary differential equations.  The downside to this choice is that it leads to a difficult coupling of the resulting ODEs that is not conducive to numerical integrations.  Instead we will approach the Euler-Lagrange equations as four second-order ODEs.

The Euler-Lagrange equation for the $u_-$ coordinates results in

\[ \frac{\partial {\mathcal{L}}}{\partial u_-} = 0 = \frac{d}{ds} \left(\frac{\partial {\mathcal{L}}}{\partial \dot u_-}\right) \]

\noindent which we write as

\begin{equation} Q_{u_-} = T_3 \ddot r + T_5 \ddot \phi_- + 2 T_6 \ddot u_- \label{ueqn} \end{equation}

\noindent for

\begin{equation} Q_{u_-} = -T_{5r} \dot r \dot \phi_- - T_{5\theta} \dot\theta \dot \phi_- - 2 T_{6r} \dot r \dot u_- - 2T_{6\theta} \dot\theta\dot u_- \label{Qu} . \end{equation}

\noindent In Eqn.~\ref{Qu} and subsequent equations, subscripts with respect to coordinates $r$ and $\theta$ represent partial derivatives of those functions with respect to the coordinate.  Similar examination of the Euler-Lagrange equations for the $\phi_-$ and $r$ coordinates results in

\begin{eqnarray} Q_{\phi_-} & =& T_2 \ddot r + 2 T_4 \ddot \phi_- + T_5 \ddot u_- \label{phieqn} \\ Q_r & =& T_2 \ddot \phi_- + T_3 \ddot u_- \label{reqn} \end{eqnarray}

\noindent where we introduce the functions

\begin{eqnarray} Q_{\phi_-} & = &  -T_{2\theta} \dot r \dot \theta - 2 T_{4r} \dot r\dot \phi_- - 2T_{4\theta} \dot \theta \dot\phi_- - T_{5r} \dot r \dot u_- - T_{5\theta} \dot\theta \dot u_- \label{Qphi} \\ Q_r & = & -T_{2\theta} \dot\theta \dot \phi_- + T_{1r} \dot\theta^2 + T_{4r} \dot\phi_-^2 + T_{5r} \dot \phi_- \dot u_- + T_{6r} \dot u_-^2. \label{Qr} \end{eqnarray}

Eqns.~\ref{ueqn}, \ref{phieqn}, and \ref{reqn} are three coupled, second-order ordinary differential equations for $\ddot u_-$, $\ddot \phi_-$ and $\ddot r$.  We algebraically uncouple these equations by first solving Eqn.~\ref{reqn} for $\ddot u_-$:

\begin{equation} \ddot u_- = \frac{1}{T_3} \left( Q_r - T_2 \ddot \phi_- \right). \end{equation}

\noindent  Likewise, Eqn.~\ref{ueqn} can be solved for $\ddot r$, replacing $\ddot u_-$ from the equation above, resulting in

\begin{equation} \ddot r = \frac{1}{T_3}\left( Q_{u_-} - T_5 \ddot \phi_- - 2\frac{T_6}{T_3} Q_r  +2 \frac{T_2 T_6}{T_3} \ddot \phi_- \right). \end{equation}

\noindent Both of these equations can then be substituted into Eqn.~\ref{phieqn}, with the resulting equation solvable for $\ddot \phi_-$.  The remaining equations then simply decouple. Defining $R_1$ and $R_2$ as

\begin{eqnarray} R_1 &=& 2 \left(T_4 + \frac{T_2^2 T_6}{T_3^2} - \frac{T_2 T_5}{T_3} \right)  \nonumber \\ R_2 &=& Q_{\phi_-}
    - \frac{T_2}{T_3} Q_{u_-} + 2 \frac{T_2 T_6}{T_3^2}  Q_r  - \frac{T_5}{T_3} Q_r \label{Rs}, \end{eqnarray}

\noindent we obtain

\begin{eqnarray} \ddot \phi_- &=& \frac{R_2}{R_1} \label{ddotphi} \\ \ddot r & = & \frac{1}{T_3} \left\{ Q_{u_-} - T_5 \left( \frac{R_2}{R_1} \right) - 2\frac{T_6}{T_3} Q_r + 2\frac{T_2 T_6}{T_3} \left( \frac{R_2}{R_1} \right) \right\} \label{ddotr} \\ \ddot u_- & = & \frac{1}{T_3} \left\{ Q_r - T_2 \left( \frac{R_2}{R_1} \right) \right\}. \label{ddotu} \end{eqnarray}

\noindent To these three equations we add to the $\theta$ Euler-Lagrange equation:

\begin{equation} \ddot \theta = \frac{1}{2T_1} \left( -T_{1\theta} \dot \theta^2 - 2 T_{1r} \dot r \dot \theta + 2 T_{2\theta} \dot r\dot\phi_- + T_{4\theta} \dot \phi_-^2 + T_{5\theta} \dot \phi_- \dot u_- + T_{6\theta} \dot u_-^2 \right). \label{ddottheta} \end{equation}

We note that these Euler-Lagrange equations did not involve first integrals of the motion, and as a result are second-order ordinary differential equations.  Since we will be computing the solution to Eqns.~\ref{ddotphi}-\ref{ddottheta} numerically, we think of them as eight first-order differential equations.  The choice of coordinates allows us to remove singularities in the differential equations due to the coordinate singularity at the event horizon.  From a computational stand-point, integrating eight first-order ordinary differential equations that lack singular terms and turning points where signs switch is well worth the disadvantage of the ugly algebra and messy looking functions that result from not utilizing the first integrals of motion.

Ultimately, we are interested in examining wave fronts of constant coordinate time $t$, which we may think of as the wave fronts constructed either by an observer at infinity or those constructed by an observer who maintains his position in the space-time (up to a scale given by the metric term $g_{rr}$ in the Boyer-Lindquist coordinates. In addition, we will project the spatial location of points along the wave front into cartesian 3-space for visualization, requiring us to keep track of the azimuthal angle $\phi$ in which the true rotational features of the metric become apparent on the wave front.  Therefore, we add two first-order ordinary differential equations for the $t$ and $\phi$ coordinates based on Eqns.~\ref{uminusdef} and \ref{phiminusdef}.  All together, we will integrate ten first-order, ordinary differential equations:

\begin{eqnarray} \dot u_-  & = & v_{u_-} \label{udot} \\ \dot r &=& v_r \\ \dot \theta  &=& v_\theta \\ \dot \phi_- &=& v_{\phi_-} \\ \dot v_{u_-} & = & \frac{1}{T_3} \left\{ Q_r - T_2 \left( \frac{R_2}{R_1} \right) \right\} \\
\dot v_r & = & \frac{1}{T_3} \left\{ Q_{u_-} - T_5 \left( \frac{R_2}{R_1} \right) - 2\frac{T_6}{T_3} Q_r + 2\frac{T_2 T_6}{T_3} \left( \frac{R_2}{R_1} \right) \right\}\\
\dot v_\theta   & = &   \frac{1}{2T_1} \left( -T_{1\theta} v_\theta^2 - 2 T_{1r} v_r v_\theta + 2 T_{2\theta} v_r v_{\phi_-}  \right. \nonumber \\ & & ~~ \left. + T_{4\theta} v_{\phi_-}^2 + T_{5\theta} v_{\phi_-} v_{u_-} + T_{6\theta} v_{u_-}^2 \right) \\
\dot v_{\phi_-} & = & \frac{R_2}{R_1} \label{phiminusdot}
\\ \dot t & = & v_{u_-} + \frac{(r^2 + a^2)}{\Delta} v_r \label{tdot} \\ \dot \phi &= &v_{\phi_-} + \frac{a}{\Delta} v_r \label{phidot}. \end{eqnarray}

Equations \ref{udot}-\ref{phiminusdot} are non-singular, while Eqns.~\ref{tdot} and \ref{phidot} are singular at the event horizon.  We will employ error-correcting adaptive step-size methods that accurately track the trajectory of the light ray through the configuration space in the the non-singular coordinates.  Error in the singular equations, Eqns.~\ref{tdot} and \ref{phidot}, will not be tracked.  In Section~\ref{rays:sec}, we further address the issue the accuracy of our simulations and the choice to track error in the Kerr coordinates only.

\subsection{Initial Conditions for Null Geodesic Wave Fronts}

We are interested in examining the wave fronts of null geodesics received by an observer at a given position and time.  Therefore, our null wave front is a past-directed, two-parameter family consisting of all the light rays received at a given initial space-time event.  The goal of this subsection is to develop the initial conditions for the light rays such that the two parameters that control positions along the wave front are made evident.  (In this section we explicitly use Greek indices to denote four-dimensional tensors and $i,j$ to denote three-dimensional spatial sub-spaces.)

We briefly return to the Boyer-Lindquist coordinates, where the space-time splitting of the metric is easiest to see.   We note that the non-zero components of the metric tensor $g_{\alpha\beta}$ are

\begin{eqnarray} g_{tt} &=& -1 + \frac{2mr}{\rho^2} \label{g00} \\ g_{t\phi} & = & -\frac{2\,m\,a\,r\, \sin^2 \theta}{\rho^2} \label{g03} \\ g_{rr} &=& \frac{\rho^2}{\Delta} \label{g11} \\ g_{\theta\theta} &=& \rho^2 \\ g_{\phi\phi} &=& (r^2 + a^2) \sin^2 \theta + \frac{2\, m\, a^2 \, r}{\rho^2} \sin^4 \theta   \label{g33}. \end{eqnarray}

\noindent Because we are working with light rays, we have a null vector $\ell^\alpha = (v_t,v_r, v_\theta, v_\phi)$ that satisfies

\[ \mathcal{L} = \frac{1}{2} g_{\alpha\beta} \ell^\alpha \ell^\beta = 0 \]

\noindent which we can solve for $v_r^2$, obtaining the relation

\begin{equation} \frac{\rho^2}{\Delta} v_r^2 = -g_{tt} v_t^2 - 2 g_{t\phi} v_t v_\phi - g_{\theta\theta} v_\theta^2 - g_{\phi\phi} v_\phi^2 \label{vr1}. \end{equation}

\begin{figure}
\begin{center}
\scalebox{1.0}{\includegraphics{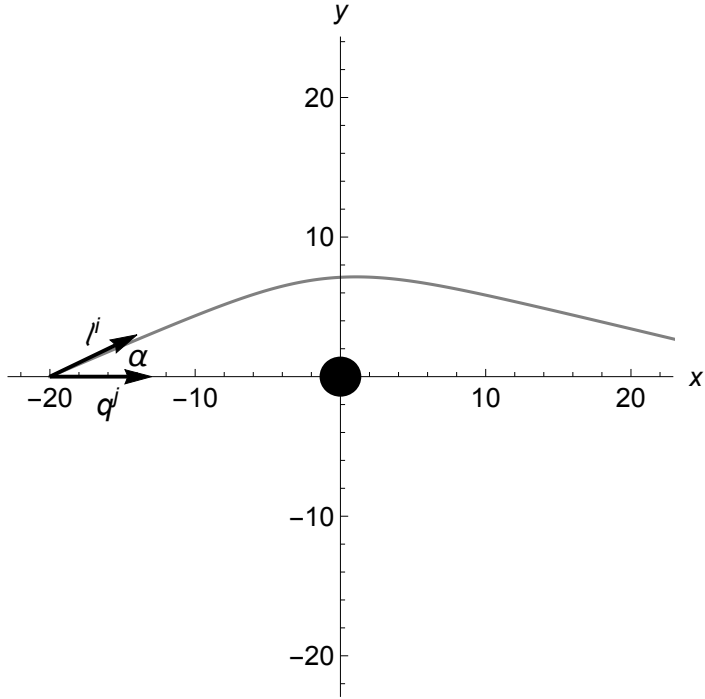}} \caption{\label{alphadef:fig}
A light ray in the equatorial plane originating at $r=20$ and $\phi=\pi$ along with two spatial vectors $\ell^i$, tangent to the spatial part of the ray, and $q^j$ which points towards the origin.  The angle $\alpha$ is defined by the inner product of these two spatial vectors, allowing us to create rings of constant opening $\alpha$ at the initial location as one of two parameters of null wave fronts. } \end{center}\end{figure}

At the initial point $(t_0, r_0, \theta_0, \phi_0)$ we define a vector that points towards the origin $q^\alpha = (0,-1,0,0)$ with spatial part $q^i = (-1,0,0)$.  We use the spatial portion of the metric, $g_{ij}$, to define an angle $\alpha$ between the spatial part of the null vector $\ell^i = (v_r, v_\theta, v_\phi)$ and $q^i$ according to the rule

\begin{equation} g_{ij} q^i \ell^j = |q| \, |\ell | \, \cos \alpha \label{dotprod}, \end{equation}

\noindent where the magnitudes are defined by $|q| = \sqrt{ g_{ij} q^i q^j}$ and $| \ell | =  \sqrt{ g_{ij} \ell^i \ell^j}$. This geometry is shown in Fig.~\ref{alphadef:fig}. The expressions in Eq.~\ref{dotprod} work out to be

\begin{eqnarray}  g_{ij} q^i q^j &=& \frac{\rho^2}{\Delta} \label{magq} \\  g_{ij} \ell^i \ell^j &=& \frac{\rho^2}{\Delta} v_r^2 + \rho^2 v_\theta^2 + g_{\phi\phi} v_\phi^2 \label{ellmag} \\ g_{ij} q^i \ell^j &=& -\frac{\rho^2}{\Delta}v_r \label{prod} .  \end{eqnarray}

Substituting these three relations into Eq.~\ref{dotprod} and squaring both sides, we obtain

\begin{equation} \frac{\rho^4}{\Delta^2} v_r^2 = \frac{\rho^2}{\Delta} \left( \frac{\rho^2}{\Delta} v_r^2 + \rho^2 v_\theta^2  + g_{\phi\phi} v_\phi^2 \right) \cos^2 \alpha. \end{equation}

\noindent Algebraically, we rearrange this expression to get

\begin{equation} \frac{\rho^2}{\Delta} v_r^2 \sin^2 \alpha = (\rho^2 v_\theta^2 + g_{\phi\phi} v_\phi^2) \cos^2 \alpha \label{mid}. \end{equation}

\noindent We can now substitute the expression we obtained for $v_r^2$ from solving $\mathcal{L} = 0$ for $v_r$ in Eq.~\ref{vr1} into Eq.~\ref{mid} which results in an expression that is written entirely in terms of $(\alpha, v_t, v_\theta, v_\phi)$ along with the coordinates:

\begin{equation} \sin^2 \alpha (-g_{tt} v_t^2 - 2g_{t\phi} v_t v_\phi - g_{\theta\theta} v_\theta^2 - g_{\phi\phi} v_\phi^2 ) = \cos^2 \alpha (\rho^2 v_\theta^2 + g_{\phi\phi} v_\phi^2) \label{mid1}. \end{equation}

\noindent Eq.~\ref{mid1} is solvable for $v_\theta$:

\begin{equation} \rho^2 v_\theta^2  = - g_{tt} (\sin^2 \alpha ) v_t^2 -  2 g_{t\phi} (\sin^2 \alpha) v_t v_\phi - g_{\phi\phi} v_\phi^2 \label{vthetalim}. \end{equation}

\noindent Since the left hand side of Eq.~\ref{vthetalim} must be positive, we have a restriction on the values of $v_\phi$.  The right hand side is quadratic in $v_\phi$ so we have as the limiting conditions the $v_\phi$ must fall in the range ${v_\phi}_a < v_\phi < {v_\phi}_b$ where we have

\begin{equation} {v_\phi}_{ab}  = \frac{-v_t}{g_{\phi\phi}} \left( g_{t\phi} \sin^2 \alpha \pm  \sin\alpha \sqrt{ g_{t\phi}^2 \sin^2 \alpha - g_{\phi\phi} g_{tt} } \right) \label{vphilim} \end{equation}

\noindent where ${v_\phi}_a$ corresponds to the minus sign in Eq.~\ref{vphilim}.  We see that there are no limits on $v_t$, so we can set $v_t = -1$ freely for past-time wave fronts.  The term in the square root of Eqn.~\ref{vphilim} is manifestly positive outside the event horizon for all angles $\alpha$.

Therefore, our procedure for creating a past-directed, two-parameter family of light rays, or a wave front, received at a single point in space-time is to pick initial coordinate locations $(t=0, r= r_o, \theta=\theta_o, \phi=\phi_o)$ with $u_- = 0$ and $\phi_- = \phi_o$, and then set $v_t = -1$.  We then set a value of $\alpha$ as our first free parameter, where $\alpha$ determines the angular opening of a ring at the initial wave front location.  The initial value of $v_\phi$ is then picked as a second parameter in the range ${v_\phi}_a < v_\phi < {v_\phi}_b$.  The initial value (up to sign) is set for $v_\theta$ using Eq.~\ref{vthetalim}, and the initial value of $v_r$ is set to ensure that the ray is a light ray as in Eq.~\ref{vr1}.  Finally, we set the initial conditions for $v_{u_-}$ and $v_{\phi_-}$ based on the initial values of the other coordinates using

\[ v_{u_-} = v_t - \frac{r^2 + a^2}{\Delta} v_r, \]
\[ v_{\phi_-}  = v_\phi - \frac{a}{\Delta}v_r, \]

\noindent defined at the initial position, $r=r_o$.

\section{Numerical Resolution and Individual Light Rays} \label{rays:sec}

To examine the properties of our numerical integration scheme, we look to light rays that pass near the black hole event horizon at

\[ r_+ = m + \sqrt{m^2 - a^2} \]

\noindent where the term $\Delta = r^2 - 2mr + a^2$ goes to zero, resulting in a coordinate singularity in the metric in Boyer-Lindquist coordinates.  If one were to chose to use Boyer-Lindquist coordinates with an adaptive step-size approach for integrating the null geodesics, these coordinate singularities would cause the estimate of the error in a given step to become very large near the event horizon, effectively decreasing the step-size to zero and stalling the light ray.  This happens even though we know that the light ray can fall into the black hole, and would cross the event horizon in a finite value of the affine parameter.  Therefore, testing the numerical integration of past directed null geodesics in the $(r, \theta, \phi_-, u_-)$ coordinates with adaptive step-size near $r_+$ is a good way to determine the effectiveness of the approach.

As a first test, we find that past-directed null geodesics directed towards the black hole can enter the event horizon in a finite number of steps with no appreciable accumulation of error. For example, we consider a past-directed ray approaching an $m=1$, $a = 0.9$ Kerr black hole that begins at $r_o = 20$ and is directed straight at the black hole in the equatorial plane with initial $\alpha = 0$.  For an  initial step-size of $h=0.0010$, this null geodesic enters the black hole in $144$ steps and accumulates no more than $6.2 \times 10^{-9}$ in total error in any coordinate.  (The error estimate here is given by Runge-Kutta-Fehlberg 4-5 error estimate, which is computed by comparing two equivalent numerical steps \cite{nrc}.)

\begin{figure}
\begin{center}
\scalebox{1.0}{\includegraphics{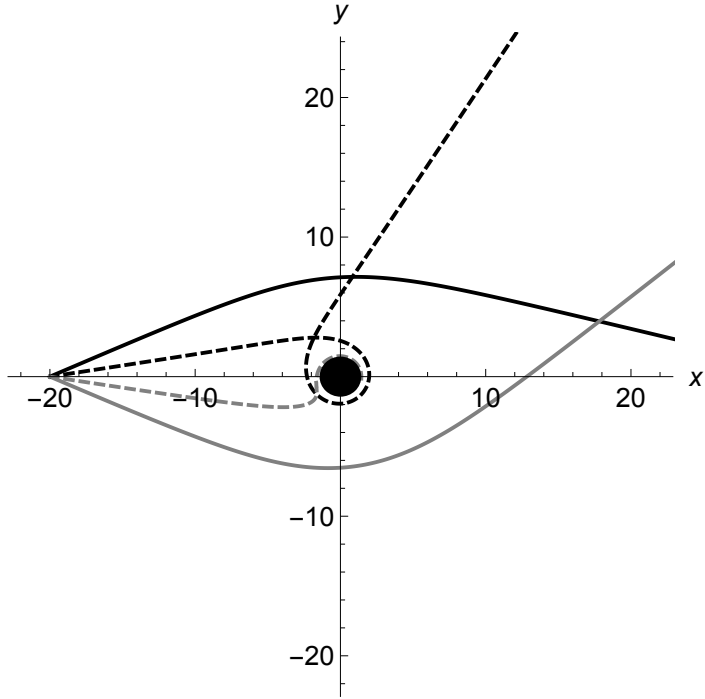}} \caption{\label{fourrays:fig}
Four light rays in the equatorial plane with initial angles $\alpha = 0.4$ and initial $v_\phi = {v_{\phi}}_b$ (solid black), $\alpha = 0.4$ and initial $v_\phi = {v_{\phi}}_a$ (solid grey), $\alpha = 0.15$ and initial $v_\phi = {v_{\phi}}_b$ (dashed black), and $\alpha = 0.15$ and initial $v_\phi = {v_{\phi}}_a$ (dashed grey) where ${v_{\phi}}_{ab}$ are defined as in Eq.~\ref{vphilim}.  The black hole has $m=1$ and $a=0.9$ and the event horizon at $r_+$ is drawn as a black circle.  The rotation of the black hole  leads to an asymmetry in the paths of the light rays.} \end{center}\end{figure}

Second, we examine the spatial paths of four past directed null geodesics in the equatorial plane for a Kerr black hole with $m=1$ and $a=0.9$.  These spatial paths are plotted according to $x = r \cos \phi$ and $y = r \sin \phi$, using the Boyer-Lindquist $\phi$ angle.  In Fig.\ref{fourrays:fig}, we see four rays, with symmetric initial opening angles for $\alpha = 0.4$ and $\alpha =0.15$, with initial $v_\phi = {v_\phi}_a$ and $v_\phi = {v_\phi}_b$ defined by Eq.~\ref{vphilim}. At the larger $\alpha$ values, the rays pass by the black hole at a relatively large distance, but clearly affected by the rotation of the black hole as they do not intersect along the axis connecting the initial point and the black hole.  For the two inner rays, we see that one ray wraps around the black hole before escaping, while its symmetric counterpart is caught up by the black hole rotation and ultimately enters the even horizon.  In fact, the ray with $\alpha = +0.15$ and initial $v_\phi = {v_\phi}_a$ exhibits a frame dragging effect and begins to rotate with the black hole, seeming to change direction near the event horizon.

\begin{figure}
\begin{center}
\scalebox{1.0}{\includegraphics{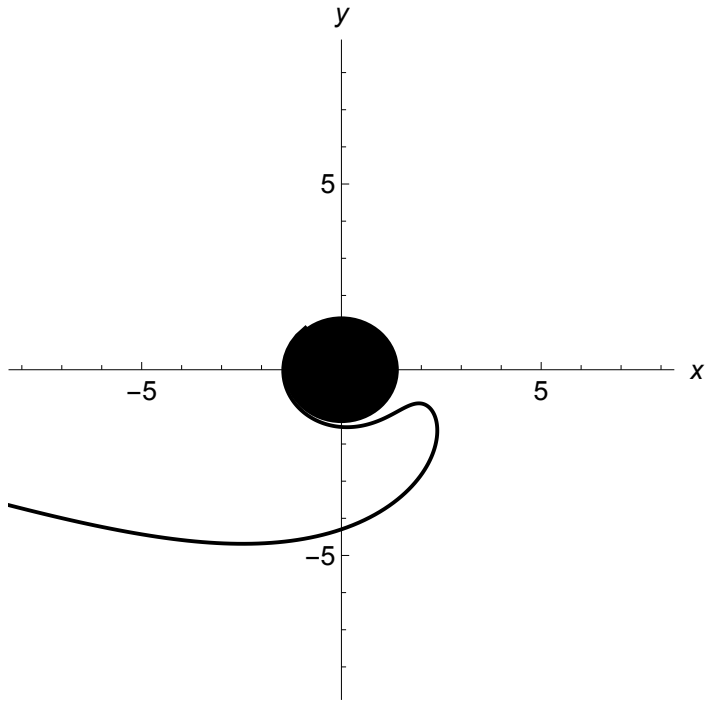}} \caption{\label{retrograde:fig}
A past directed null geodesic in the equatorial plane of a Kerr black hole with $m=1$ and $a=0.9$ that shows a null geodesic that changes direction due to the clockwise rotation of the space-time.} \end{center}\end{figure}

In making these plots, it became apparent that the adaptive step-size approach, which continues to take large steps if no significant error is found in the $(r,v_r, \theta, v_\theta, \phi_-, v_{\phi_-}, u_-, v_{u_-})$ coordinates, runs into trouble plotting points using the true $\phi$ angle.  Since the differential equations governing the $t$ and $\phi$ coordinates, Eqns.~\ref{tdot} and \ref{phidot}, are singular at the event horizon, even though we achieve highly accurate steps in the new coordinates, near the event horizon the $\phi$ and $t$ coordinates take large jumps.  For this reason, we have modified our adaptive step-size code to require small, fixed-sized steps near $r_+$.  With this modification to the step-size algorithm, we can achieve smooth plots in the spatial position which are accurate to one part in $10^9$ or better in the coordinates that extend across the event horizon.  Figure~\ref{retrograde:fig} shows a smooth plot of a ray that comes very close to the event horizon before reversing direction due to the rotation of the black hole and entering into the event horizon.

\begin{figure}
\begin{center}
\scalebox{1.0}{\includegraphics{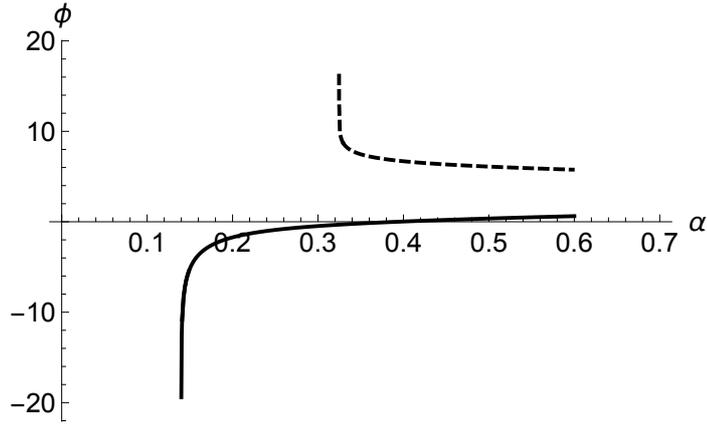}} \caption{\label{alphaplot:fig}
A plot of the $\phi$ angle at $r=30$ for rays that have escaped a Kerr black hole with $m=1$ and $a=0.9$ for null geodesics approaching the black hole and originating at $r=20$ and $\phi = \pi$ and remaining in the equatorial plane.  On the horizontal axis is the initial angle $\alpha$ between the null geodesic and the vector pointing towards the origin. The final $\phi$ angle for rays with initial $v_\phi = {v_\phi}_a$ are plotted in solid black, while the final $\phi$ angle for rays with initial ${v_\phi}_b$ are dashed.  The asymmetry in the final $\phi$ values for a given $\alpha$ is due to the rotation of the black hole.  The plot shows that our numerical code is able to track the $\phi$ coordinates for rays that wrap around the black hole about three times in either direction.} \end{center}\end{figure}

Finally in Fig.~\ref{alphaplot:fig}, we plot the $\phi$ coordinate as a function of the initial angle $\alpha$ for equatorial light rays when the ray has reached $r=30$ after escaping the black hole. The plot includes rays with a given $\alpha$ value in the equatorial plane for both signs in Eq.~\ref{vphilim}. We see that we can achieve highly accurate plots that wrap many times around the black hole achieving final phi values close to 20 radians.  There is an asymmetry present between the rays with ${v_\phi}_a$ and ${v_\phi}_b$ due to the black hole rotation.

\section{The Origin of Astroidal Caustics} \label{astroidal:sec}

Rauch and Blandford \cite{rauch} first reported that the caustic surfaces of the Kerr metric are tubes with non-symmetric astroidal cross sections; a result that was later confirmed by multiple authors \cite{sereno2008, bozza2008, grould}.  Caustic surfaces are the locations in space-time of a vanishing geodesic deviation vector where light rays have focused.  Previous papers on the caustics in the Kerr metric computed the location of the caustics by taking numerical derivatives between light rays.

Caustic surfaces are also the spatial locations of wave front singularities. As the wave front moves in time, the position of the singularities along the wave front traces out the caustic shape.  The central purpose of this paper is to examine how the astroidal shaped caustic tube arises in the Kerr metric from the wave front perspective.

Astroidal caustics are well understood to arise from tracing the inward directed wave front of an ellipse.  Continuously shining lights directed perpendicular from the inner surface of an ellipse focus in the astroidal shape of four connected cusps.

From the wave front perspective, as the elliptical wave front collapses, it forms cusp singularities.  The location of the cusps move, tracing out the astroidal shape.  Cusp singularities are one of the canonical forms of wave front singularities that are stable under perturbations.  All astroidal shaped caustics result from the evolution of wave fronts with cusps, a result well explained in work by V.I. Arnol'd \cite{arnold1} and others.

Therefore, for an astroidal tube to form as the caustic of point sources of light in the Kerr metric, we expect to see the collapse of some sort of tube.  The rotational properties of the Kerr metric would be expected to collapse such a tube non-symmetrically, leading to cusps in the cross section of the wave fronts.

\section{Schwarzschild Wave Fronts} \label{schwarz:sec}

In the $a=0$ case, our equations reduce to the null geodesic equations of the Schwarzschild metric.  The wave fronts of the Schwarzschild metric must be axially symmetric about the line connecting the origin and the initial point light source.  Because we are plotting wave fronts of constant coordinate time (or the time coordinate of an observer at infinity), the wave fronts hang up along the event horizon of the black hole, which in all figures is shown in solid black for reference.

We choose initial conditions with $m=1$, and initial $r = 20$ with $\phi_- = \phi = \pi$ and $\theta = \pi/2$.  This places the initial point on the $-\hat x$ axis.  We choose to show only the portion of the wave front that approaches the black hole.  The remainder of the wave front smoothly connects around this portion.

\begin{figure}
\begin{center}
\scalebox{0.6}{\includegraphics{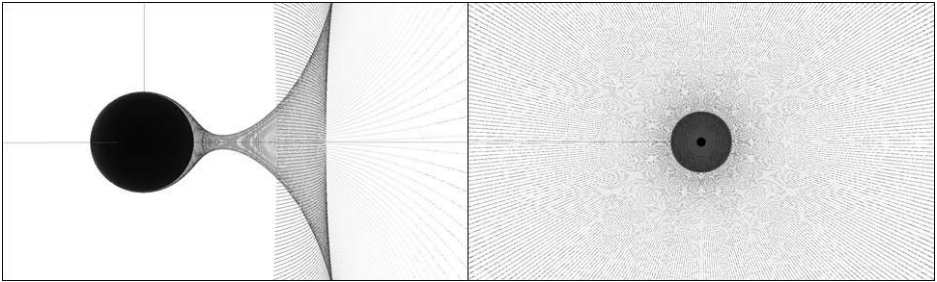}} \caption{\label{schwarz336:fig}
Two views of a portion of the null wave front in the Schwarzschild space time for a flash of point light source located at $r=20$, $\phi = \pi$ and $\theta = \pi/2$ at $t=-33.6$.  In the left panel, we see the formation of a tube symmetric with the axis connecting the initial point and the origin from the side where the $+\hat x$ axis is directed to the right.  The right hand panel looks directly down the $+\hat x$ axis, showing that the tube's opening is symmetric with the axis of symmetry.} \end{center}\end{figure}

\begin{figure}
\begin{center}
\scalebox{0.9}{\includegraphics{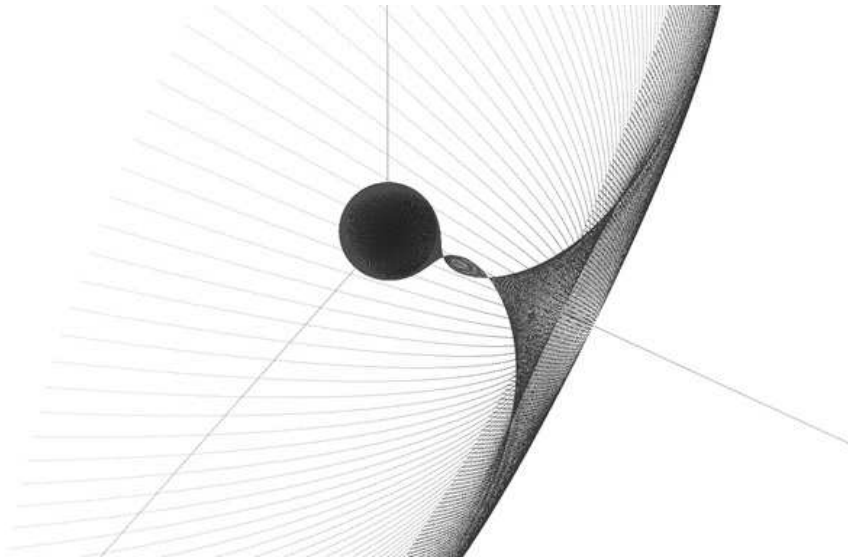}} \caption{\label{schwarz345:fig}
The null wave front in the Schwarzschild space time for a flash of a point light source located at $r=20$, $\phi = \pi$ and $\theta = \pi/2$ at $t=-34.5$.  The tube has collapsed symmetrically at two twist points, so that the wave front singularities are two cross-over points without structure.  The $\hat x$ axis of symmetry points towards the lower right.} \end{center}\end{figure}

Figure~\ref{schwarz336:fig} shows the wave front at $t= -33.6$ as it passes the black hole.  We see that a tube is formed which is symmetric with the axis connecting the initial light ray point and the origin which we take to the $\hat x$ axis. This tube collapses first at a single point along the $\hat x$ axis and then forms two singular points, one moving away from the black hole and the other moving towards the black hole as the wave front advances, as is shown in Fig.~\ref{schwarz345:fig} at $t=-34.5$.  Part of the wave front corresponds to light rays that are circling the black hole back towards the initial location, and so as the wave front advances, a new tube forms on the side closest to the initial location and the process repeats.

Hence, in the Schwarzschild case, the caustic surface is a line stretching along the axis of symmetry connecting the initial location and the origin.  The wave front singularities are simple points where an entire ring of light rays collapses simultaneously.

\section{Kerr Wave Fronts from Point Sources in the Equatorial Plane} \label{wavefronts:sec}

All non-zero spin parameter wave fronts will exhibit the same generic features, but the features are easier to resolve for large spin parameter $a$.  We begin by considering wave fronts from point sources at initial $r = 20$ and $a = 0.9$ for $m = 1$ with $\phi_- = \phi = \pi$ so that the initial point is along the $-\hat x$ axis in the azimuthal plane.  As in the Schwarzschild case, we focus on the portion of the wave front that approaches and passes by the black hole.

\begin{figure}
\begin{center}
\scalebox{0.2}{\includegraphics{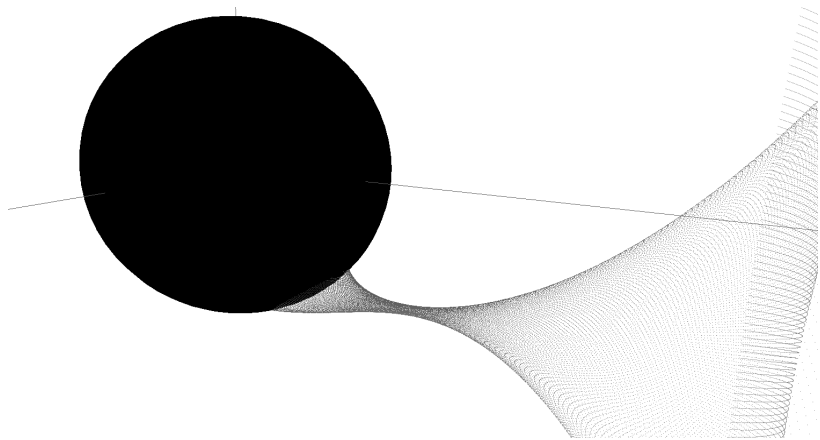}} \caption{\label{kerr336:fig}
The null wave front in the Kerr space time with $a=0.9$ for a flash of a point light source located at $r=20$, $\phi = \pi$ and $\theta = \pi/2$ at $t=-33.6$. We see that the wave front has formed a tube, as in the Schwarzschild case, but that this tube is no longer symmetric with the $\hat x$ axis.  In the figure, the $\hat x$ axis points to the right.} \end{center}\end{figure}

\begin{figure}
\begin{center}
\scalebox{0.2}{\includegraphics{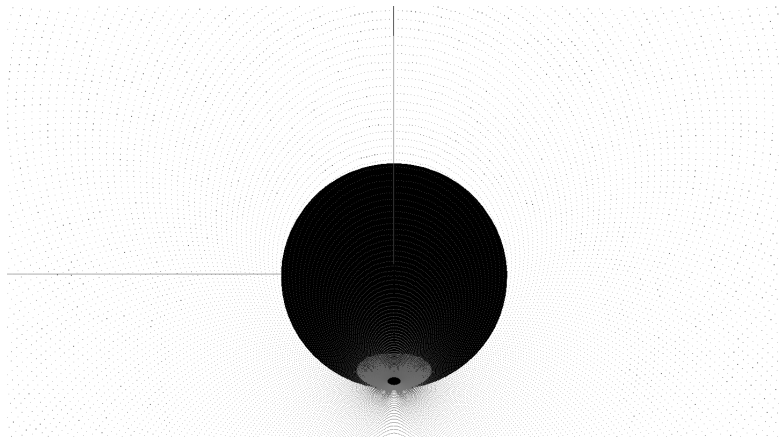}} \caption{\label{kerr336_front:fig}
The same wave front as in Fig.~\ref{kerr336:fig} viewed from a position in front of the wave front with the $\hat x$ axis pointing towards the viewer and slightly towards the top of the page and the $\hat y$ axis pointing to the left.  We see that the opening of the tube is not symmetric as was the case in the Schwarzschild wave front.  Also, there is sculpting along the outer portion of the wave front where the opening of the tube is forming a mound that points towards the $\hat x$ axis as the tube closes along the center bottom of this figure.} \end{center}\end{figure}

Similar to the Schwarschild case, we see that the time delay effect near the event horizon of the Kerr black hole leads to the formation of a tube in the wave front.  However, in Figure~\ref{kerr336:fig} we see at $t=-33.6$ that the opening of the wave front has been displaced from the $+\hat x$ in the direction towards the $-\hat y$ axis due to the frame dragging effect of the black hole.  The tube portion of the wave front is collapsing non-symmetrically, forming a ridge displaced in the azimuthal plane from the axis connecting the observer and the origin in the direction of the rotation, as shown in Fig.~\ref{kerr336_front:fig}.

\begin{figure}
\begin{center}
\scalebox{0.2}{\includegraphics{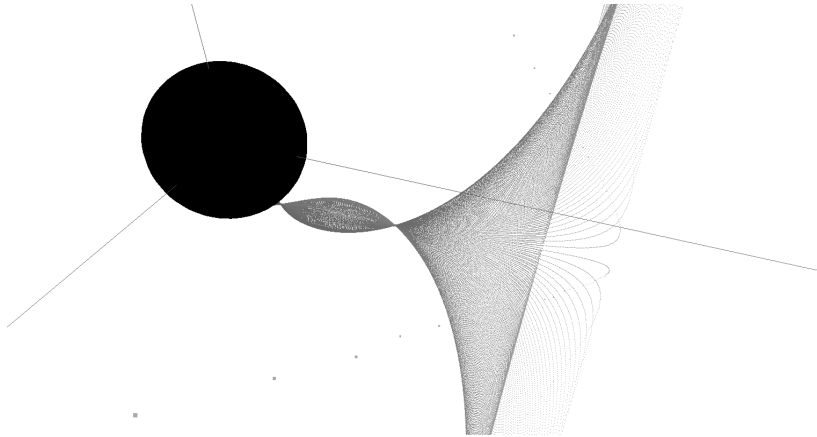}} \caption{\label{kerr345:fig}
The null wave front in the Kerr space time with $a=0.9$ from an initial point located at $r=20$, $\phi = \pi$ and $\theta = \pi/2$ at $t=-34.5$. The wave front tube from Fig.~\ref{kerr336:fig} has pulled through itself in a pair of crossings similar fashion to the Schwarzschild case.} \end{center}\end{figure}

\begin{figure}
\begin{center}
\scalebox{0.2}{\includegraphics{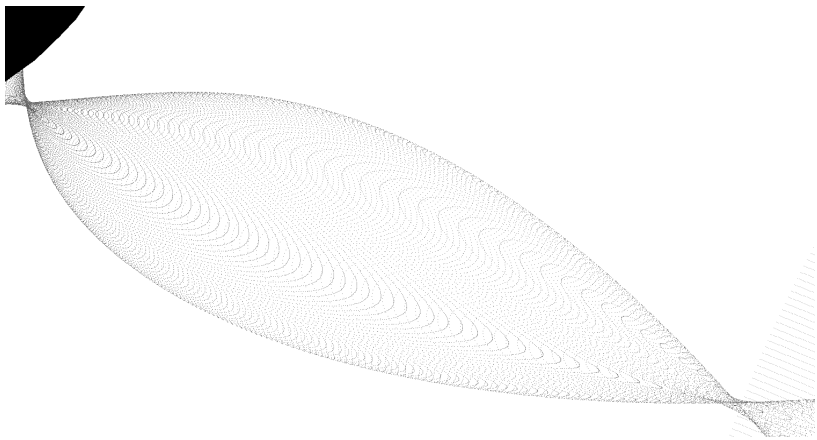}} \caption{\label{kerr345_close:fig}
A close inspection of the crossings in Fig.~\ref{kerr345:fig} reveals that there is internal structure and that the crossings are not simple twist points in the Kerr wave front at $t = -34.5$.  The asymmetry of the tube in Fig.~\ref{kerr336:fig} has led to a wave front where neighboring rays are crossing at different locations.} \end{center}\end{figure}

Figure~\ref{kerr345:fig} shows that as the wave front's tube fully collapses, at $t=-34.5$, the collapse has not occurred at single twist points, but instead along lines.  This central structure pulls through itself, and part of the wave front wraps around the black hole repeating the process on the side closer to the original flash point.  Figure~\ref{kerr345_close:fig} shows the same wave front, close to the crossings.  Unlike the Schwarzschild case, there is clear structure to the overlap of the tube.

\begin{figure}
\begin{center}
\scalebox{0.2}{\includegraphics{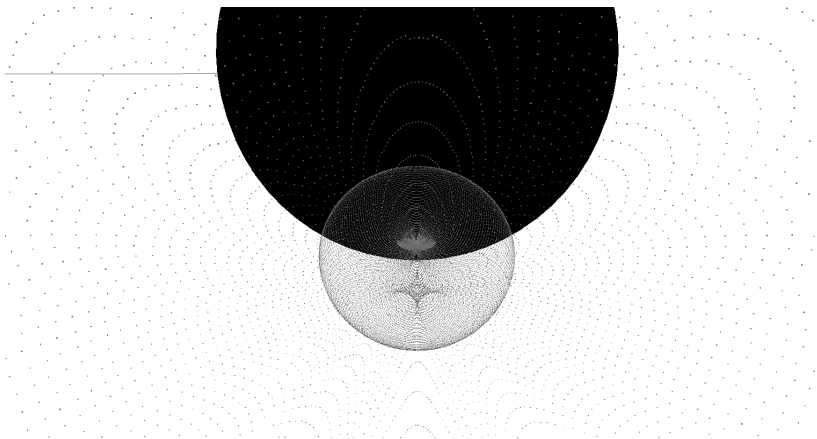}} \caption{\label{kerr345_astroid1:fig}
The null wave front in the Kerr space time with $a=0.9$ for an initial point located at $r=20$, $\phi = \pi$ and $\theta = \pi/2$ from a vantage point directly in line with the crossings and internal structure shown in Fig.~\ref{kerr345_close:fig}. We see an astroidal caustic has formed at the crossing point of the tube that is furthest from the black hole.} \end{center}\end{figure}

\begin{figure}
\begin{center}
\scalebox{0.2}{\includegraphics{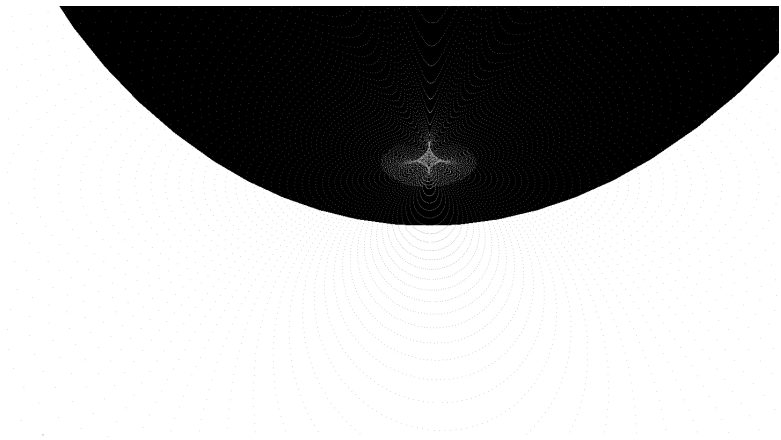}} \caption{\label{kerr345_astroid2:fig}
Moving through the tube in Fig.~\ref{kerr345_astroid1:fig} directly towards the black hole allows one to zoom in on the crossing nearer the event horizon.   We see a second astroidal caustic against the back-drop of the event horizon.} \end{center}\end{figure}

The tell-tale sign of the formation of an astroidal tube in the caustic would be the presence of cusp singularities in individual wave fronts.  These become visible by repositioning the vantage point to look directly at the crossing of the wave front at time $t=-34.5$.  Figure~\ref{kerr345_astroid1:fig} zooms in on the wave front from directly in front of the crossing closest to the outer portion of the front.  Zooming in further,  towards the black hole to examine the crossing nearer the event horizon we see a second astroid in Fig.~\ref{kerr345_astroid2:fig}.  Both figures show clear astroidal caustics. As the wave front moves with time, the positions of these astroids trace out the astroidal tubes of the literature.  What is referred to as the primary caustic is the astroidal caustic furthest from the black hole, which is moving away from the event horizon.  The secondary caustic is the astroid closer to the event horizon, which moves towards the event horizon and wraps around the black hole.

\begin{figure}
\begin{center}
\scalebox{0.2}{\includegraphics{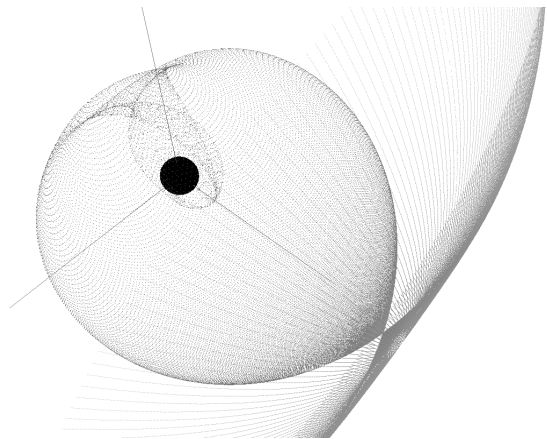}} \caption{\label{kerr55:fig}
A portion of the null wave front in the Kerr space time with $a=0.9$ and $m=1$ for a flash from a point source located at $r=20$, $\phi = \pi$ and $\theta = \pi/2$ at $t=-55$. The tube portion of the wave front seen in Fig.~\ref{kerr345:fig} has wrapped around the black hole, leading to a large astroidal caustic.} \end{center}\end{figure}

\begin{figure}
\begin{center}
\scalebox{0.6}{\includegraphics{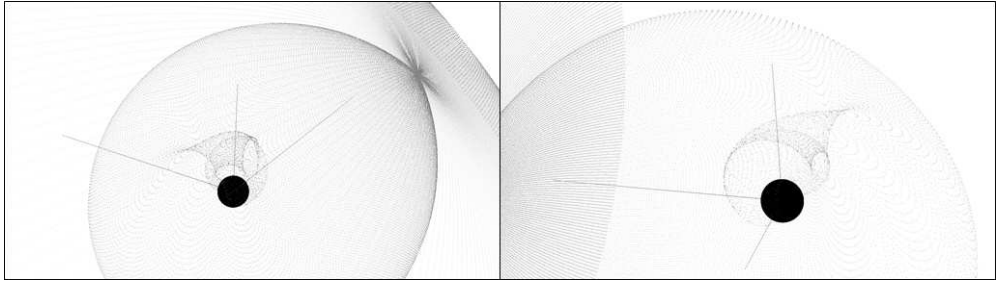}} \caption{\label{kerr55close:fig}
Two views of the null wave front in the Kerr space time with $a=0.9$ for an initial point at $r=20$, $\phi = \pi$ and $\theta = \pi/2$ at $t=-55$ near the event horizon.} \end{center}\end{figure}

Several views of the wave front at late times are provided in Figs.~\ref{kerr55:fig} and \ref{kerr55close:fig} for wave fronts with the same range in initial parameters at $t=-55.0$.  In Fig.~\ref{kerr55:fig} we see portions of the wave front have wrapped around the Kerr event horizon.  A large astroidal caustic appears in the portion of the wave front that has wrapped around the black hole.

\section{Kerr Wave Fronts Outside Plane of Symmetry} \label{outofplane:sec}

\begin{figure}
\begin{center}
\scalebox{0.2}{\includegraphics{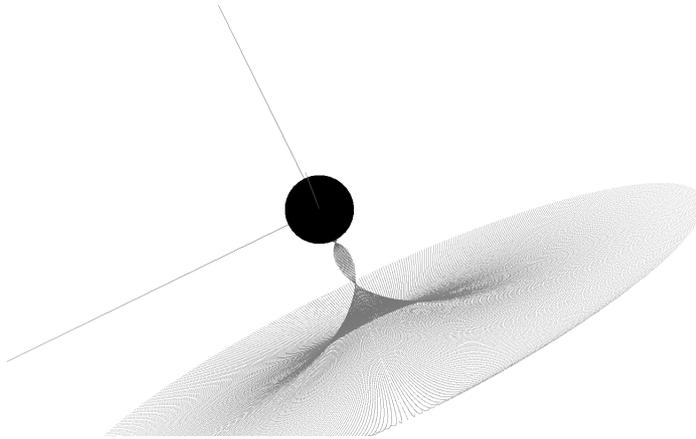}} \caption{\label{kerr_theta0:fig}
A portion of the null wave front in the Kerr space time with $a=0.9$ and $m=1$ for a flash from a point source at $r=20$, $\theta \approx 0$ at $t=-34.5$. The $\hat z$ axis points towards the upper left.  For wave fronts originating along the axis of symmetry, the wave front crosses symmetrically in twist points as in the Schwarzschild case.} \end{center}\end{figure}

Wave fronts emanating from points not in the equatorial plane can differ structurally from those with initial $\theta = \pi/2$. Figure~\ref{kerr_theta0:fig} shows the wave front from a point source on the $+\hat z$ axis in the Kerr geometry with $a=0.9$.  As expected,  the wave front in the Kerr geometry forms a tube that collapses symmetrically at two twist points, just as in the Schwarzschild case as in Fig.~\ref{schwarz345:fig}.  The frame dragging of the Kerr black hole causes the individual points on the wave front to rotate, but the caustic structure for wave fronts emanating from points along the $\hat z$ axis are individual twist points.

\begin{figure}
\begin{center}
\scalebox{0.2}{\includegraphics{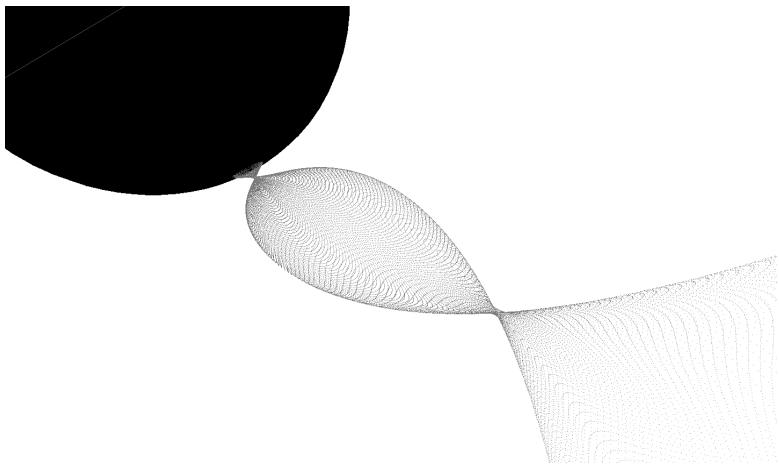}} \caption{\label{kerr_theta1.2:fig}
A portion of the null wave front in the Kerr space time with $a=0.9$ and $m=1$ for a flash of light from a point source at $r=20$ and $\theta=1.2$ for $t=-34.5$ close to the black hole.  The wave front displays similar non-symmetric crossings leading to astroidal caustics as in Fig.~\ref{kerr345_astroid1:fig}.} \end{center}\end{figure}

Figure~\ref{kerr_theta1.2:fig} shows the wave front for initial $r=20$ but $\theta = 1.2$ radians at $t = -34.5$.  We see a similar formation of a tube that collapses non-symmetrically, leading to the same general features as in the initial $\theta = \pi/2$ case. Zooming in on the wave front reveals the same basic caustic structure as before.

\section{Discussion}\label{discussion:sec}

This paper expands the literature by demonstrating how an astroidal caustic tube arises in the Kerr space-time from the wave front perspective.  Previous work alternatively has examined the caustics from the stand-point of the geodesic deviation equation \cite{rauch} and the lens mapping, for instance in \cite{sereno2008,bozza2008}.  The word ``caustic'' has slightly different meanings in these settings, and there are subtle relations between the wave front, geodesic deviation, and lens mapping approaches \cite{pettersbook}.

In the traditional presentation of gravitational lensing, the lens is treated as a weak-field perturbation of an underlying space-time, for instance a Friedmann-Robertson-Walker metric.  Then the perturbation is shown to act over a spatial depth that is small compared with the distance between the observer and sources of light behind the lens.  This justifies the introduction of a lens plane, and allows for a mapping of positions in the lens plane to the source plane in which the background sources lie.  A more general approach to the lens mapping takes image locations on (a portion of) the two-dimensional sky of the observer and maps the image locations into points in the space-time where sources lie (even if these points don't comprise a plane).

Mathematically, we can use $\vec x$ to denote positions in the lens plane, or on the observer's sky, and $\vec y$ as positions in the source plane, and denote the lens mapping as

\begin{equation} \vec y = \vec \eta (\vec x). \label{lensmap} \end{equation}

\noindent The positions where det[Jac $\vec \eta](\vec x) = 0$ trace out the critical curve in the lens plane.  The projection of those positions into the source plane via the lens mapping, Eqn.~\ref{lensmap}, are the caustics of the lens map.  The caustic curve in the source plane separates regions where sources might be located that map into different numbers of observed images, which is the cause of multiple imaging in gravitational lensing.  Points on the caustic in the source plane are also conjugate points in relation to the location of the observer: null geodesics connecting the observer and the point in the source plane on the caustic have a vanishing geodesic deviation vector at both locations.  This leads to high magnification of sources located near the caustic.

The papers by V. Bozza \cite{bozza2008} and Sereno \& DeLuca \cite{sereno2008} take the lens mapping perspective in finding the caustics.  Both papers define a lens mapping from a two parameter sky of the observer into positions in the space, and they compute locations where the determinant of the Jacobian of the mapping is zero.  The lens mappings of both authors though are slightly non-standard from the lens mapping stand-point as they don't employ a single source plane - which allows them to locate the entire caustic tube.

The original paper by Rauch and Blandford \cite{rauch} located the caustic tube by integrating the geodesic deviation equations along the null geodesics.  In that way, they located the caustic tube without formally considering the null geodesics as part of a gravitational lensing system.

As Petters et al. \cite{pettersbook} discuss, there is a relationship between the singularities of the wave front and the caustics of the lens mapping.  When a wave front is emanating from a single space-time point, the singularities of a given wave front at a particular $t$ value are part of the caustic surface which locates all the points conjugate to initial point.  The intersection of a source plane with this caustic surface provides the caustics of the lens map from the observer's sky into the source plane.  Moving that source plane fills out the caustic surface in the space-time.  Put another way, as the wave front advances, the locations of the wave front singularities trace out the caustic surface.

What is unique about the approach taken in this paper is that we are able to see the specific wave front singularities that lead to astroidal caustic tube found by Rauch \& Blandford using the geodesic deviation equations and other authors using the idea of a lens mapping.  Notably, in Fig.~\ref{kerr345_astroid1:fig} and \ref{kerr345_astroid2:fig}, we see both the primary and secondary astroidal wave front singularities in the wave front itself.

\section{Conclusion}

The Kerr metric, similar metrics, or perturbations of it, remain a test-bed for various theories of gravity or gravitational structures \cite{plasma, sen}.  Both analytic and perturbative approaches to examining light rays in these metrics continue to be studied \cite{iyer, postN, gauss}.

The primary contribution of this paper is to show that the origin of the astroidal tube-shaped caustic surface for the Kerr geometry arises from the non-symmetric collapse of the wave front.  This result provides a new insight into the results reported elsewhere in the literature that located the caustic surface through different techniques: integrating the geodesic deviation equations or finding the locations where the lensing map is singular.

Our particular methods to integrating the null geodesics differed in two ways from most of the literature.  First, we chose to integrate the null geodesics in the Kerr coordinates, $(u_-, r, \theta,\phi_-)$, and second, we chose to integrate the null geodesic equations numerically.  These choices allowed us to create numerical integrations accurate to one part in $10^9$ by utilizing adaptive step-size near the event horizon.  We believe that these methods are the most accurate numerical methods available, and that they, or a similar choice, should be used when comparing observation data to theory in future experiments.

As a result of our choices of coordinates and methods, we are able to simultaneously integrate hundreds of thousands of light rays on a standard work station, sort them by time slices, and render highly detailed still images or movies of the advancing wave fronts.  The choice of coordinates that extend across the event horizon and specially written numerical integration and visualization programs proved useful in understanding the underlying structures of the wave fronts.

The numerical methods we employ in this paper may be of use in determining the shadow of Kerr or similar black holes and considering the appearance of objects orbiting close to the event horizon, for instance in papers such as the recent paper by Cunha \& Herdeiro (2018) \cite{cunha}.  Our methods are currently designed to trace light rays received by a given observer at a given instant back in time to find the wave fronts.  To find the appearance of a black hole shadow, we would need to slightly invert our methods to project onto the observer's sky all the light rays from say evenly spaced sources.  (Our methods begin with a two-parameter family of light rays spreading out evenly across the sky.  To see a black hole shadow, one wants to begin with evenly spaced light sources that the observer views non-evenly.  This is a complicated computational task with no good, generic methods.) We will leave the modification of our code to see the black hole shadow to future studies.

\section{Appendix: Future-Directed Null Geodesics}

\begin{figure}
\begin{center}
\scalebox{1.0}{\includegraphics{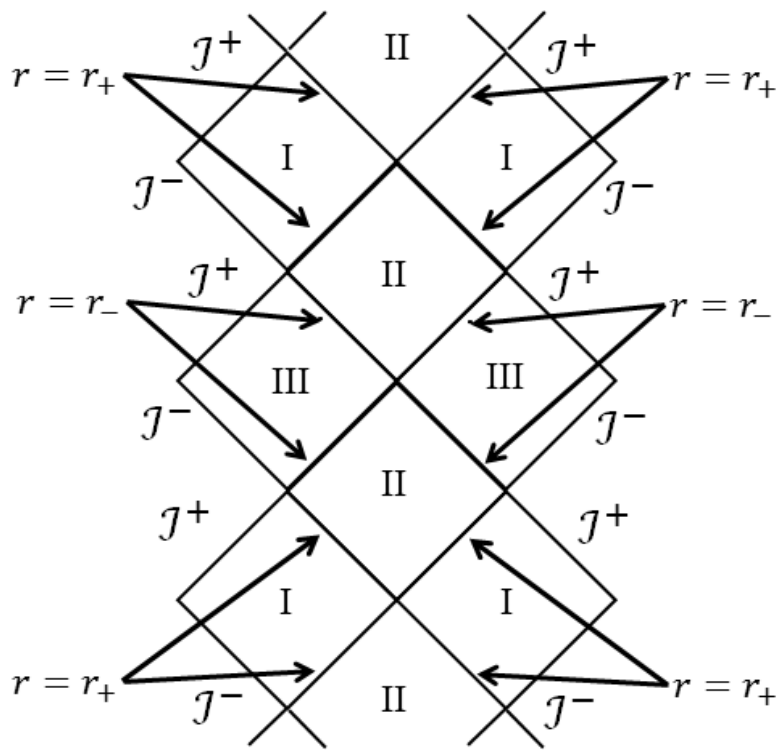}} \caption{\label{conformal:fig}
The key elements of the conformal structure of the Kerr space time as presented in Hawking \& Ellis Figure 28 \cite{HE}. The numerical methods of this paper work well to integrate a past directed light ray that starts in region I and falls into the black hole (lower region II) if the choice of coordinates is $(u_-, \phi_-)$.  To reach the black hole along future directed light rays (upper region II), the correct choice of coordinates is $(u_+, \phi_+)$.} \end{center}\end{figure}

Throughout this paper, we have been careful to indicate that the choice of coordinates $(r, \theta, \phi_-, u_-)$ was appropriate for past-directed null geodesics in use with adaptive step-size methods.  This is because the null coordinate $u_-$ is well suited for crossing the event horizon in the past-directed orientation.  If one considers the conformal structure of the Kerr solution, as represented in Fig.~\ref{conformal:fig} \cite{HE} for example, our light rays are beginning in a standard region I (for sake of argument in a region I on the right).  Past directed null geodesics in the $(r, \theta, \phi_-, u_-)$ coordinates smoothly cross the event horizon into the region II (the ``white hole,'' to the lower left) or approach past null infinity.  While one can integrate these coordinates forward in time, they are not well suited for approaching future null infinity or the copy of region II (the ``black hole,'' to the upper left).  In fact, the adaptive step-size code we implement in this paper drives the step-size to zero near the event horizon when using future-directed null geodesics in the $(r, \theta, \phi_-, u_-)$ coordinates.

To compute future-directed null geodesics, it is appropriate to change coordinates to $(r, \theta,\phi_+,u_+)$ using the coordinate transformation

\begin{eqnarray} du_+ & = &  dt + (r^2+a^2) \Delta^{-1} dr \\ d\phi_+ & = & d\phi + a \Delta^{-1} dr. \end{eqnarray}

\noindent In these coordinates, one can smoothly integrate using adaptive step-size for future-directed null geodesics, reaching future null infinity or the region II corresponding to the ``black hole''.  The only difference in the mathematics presented in this paper are that the terms in the metric in front of the $dr d\phi_-$ and $dr du_-$  change sign. This corresponds to a change in sign of our functions labeled $T_2$ and $T_3$ in the Lagrangian and subsequent Euler-Lagrange equations. It is for this reason that we did not simplify our equations by noting that $T_3=-1$ throughout Section \ref{numerical:sec}.  In our computer code, we have coded for the option of using either coordinate system, where if we want future-directed rays, we use the $u_+$ and $\phi_+$ coordinates, and if we want past-directed light rays, we use the $u_-$ and $\phi_-$ coordinates.

\begin{acknowledgements}
EG was supported on this work by an Undergraduate Research Summer Grant from the Adrian Tinsley Program at Bridgewater State University.
 KR and HW were supported under two separate NASA Space Grants from the Massachusetts Space Consortium.
\end{acknowledgements}



\end{document}